\documentclass[fleqn,10pt]{wlscirep}
\usepackage[utf8]{inputenc}
\usepackage[T1]{fontenc}
\usepackage[acronym, automake]{glossaries-extra}
\usepackage[ruled, linesnumbered]{algorithm2e}
\usepackage{amsmath, amsfonts}
\usepackage{booktabs}
\usepackage{longtable}
\usepackage{makecell}
\usepackage{xcolor}
\usepackage{colortbl}
\usepackage{graphicx}
\usepackage{multicol}
\usepackage{multirow}
\usepackage{caption}
\usepackage{subcaption}

\title{Capturing functional connectomics using Riemannian partial least squares}

\author[1,*]{Matt Ryan}
\author[1]{Gary Glonek}
\author[1]{Jono Tuke}
\author[1]{Melissa Humphries}
\affil[1]{The University of Adelaide, School of Computer and Mathematical Sciences, Adelaide, 5005, Australia}

\affil[*]{matthew.ryan@adelaide.edu.au}

\begin{abstract}

    For neurological disorders and diseases, functional and anatomical connectomes of the human brain can be used to better inform targeted interventions and treatment strategies. Functional magnetic resonance imaging (fMRI) is a non-invasive neuroimaging technique that captures spatio-temporal brain function through blood flow over time.  FMRI can be used to study the functional connectome through the functional connectivity matrix; that is, Pearson's correlation matrix between time series from the regions of interest of an fMRI image.  One approach to analysing functional connectivity is using partial least squares (PLS), a multivariate regression technique designed for high-dimensional predictor data.  However, analysing functional connectivity with PLS ignores a key property of the functional connectivity matrix; namely, these matrices are positive definite.  To account for this, we introduce a generalisation of PLS to Riemannian manifolds, called R-PLS, and apply it to symmetric positive definite matrices with the affine invariant geometry.  We apply R-PLS to two functional imaging datasets: COBRE, which investigates functional differences between schizophrenic patients and healthy controls, and; ABIDE, which compares people with autism spectrum disorder and neurotypical controls. 
 Using the variable importance in the projection statistic on the results of R-PLS, we identify key functional connections in each dataset   that are well represented in the literature. Given the generality of R-PLS, this method has potential to open up new avenues for multi-model imaging analysis linking structural and functional connectomics. 
\end{abstract}

\setabbreviationstyle[acronym]{long-short}

\newglossaryentry{voxel}{
	name = voxel,
	description = {A volumetric pixel.}
}
\newglossaryentry{TR}{
	name = repetition time,
	description = {How often scans are taken in the fMRI process.}
}

\newacronym{pls}{PLS}{partial least squares}
\newacronym{rpls}{R-PLS}{Riemannian partial least squares}
\newacronym{pca}{PCA}{principal component analysis}
\newacronym{pga}{PGA}{principal geodesic analysis}
\newacronym{pcr}{PCR}{principal component regression}
\newacronym{bold}{BOLD}{blood-oxygen-level-dependant}
\newacronym{fmri}{fMRI}{functional magnetic resonance imaging}
\newacronym[\glslongpluralkey={regions of interest}]{roi}{ROI}
{region of interest}
\newacronym{mri}{MRI}{magnetic resonance imaging}
\newacronym{mdmr}{MDMR}{multivariate distance matrix regression}
\newacronym{mds}{MDS}{multidimensional scaling}
\newacronym{spd}{SPD}{symmetric positive definite}
\newacronym{cca}{CCA}{canonical correlations analysis}
\newacronym{nipals}{NIPALS}{non-linear iterative partial least squares}
\newacronym{rnipals}{R-NIPALS}{Riemannian non-linear iterative partial least squares}
\newacronym{tnipals}{tNIPALS}{tangent non-linear iterative partial least squares}
\newacronym{aal}{AAL}{automated anatomic labelling}
\newacronym{msdl}{MSDL}{multi-subject dictionary learning}
\newacronym{rmse}{RMSE}{root mean square error}
\newacronym{asd}{ASD}{autism spectrum disorder}
\newacronym{ntc}{NTC}{neurotypical controls}
\newacronym{dmn}{DMN}{Default Mode Network}
\newacronym{van}{VAN}{Ventral Attention Network}
\newacronym{aips}{Ant IPS}{Anterior Intraparietal Sulcus}
\newacronym{manova}{MANOVA}{multivariate analysis of variance}
\newacronym{anova}{ANOVA}{analysis of variance}
\newacronym{mmr}{MMR}{multivariate multiple regression}
\newacronym{anosim}{ANOSIM}{analysis of similarities}
\newacronym{svd}{SVD}{singular value decomposition}
\newacronym{iqr}{IQR}{interquartile range}
\newacronym{vip}{VIP}{variable importance in the projection}
\makeglossaries
\newcommand{\mat}[1]{\boldsymbol{#1}}
\renewcommand{\vec}[1]{\boldsymbol{#1}}

\newcommand{\RR}{\mathbb{R}}
\DeclareMathOperator{\Exp}{Exp}
\DeclareMathOperator{\Log}{Log}
\DeclareMathOperator{\argmin}{argmin}
\DeclareMathOperator{\tr}{Tr}

\newcommand{\citep}{\cite}
\newcommand{\citet}{\cite}

\begin{document}

\flushbottom
\maketitle


%
%
\thispagestyle{empty}

\section*{Introduction}

The functional and anatomical connections of the human brain form complex networks that link the infrastructure of our minds.  Understanding these connectomes has the potential to provide insight into the effect of neurological diseases which can be used to better inform targeted interventions and treatment strategies\cite{Contreras2015, Yang2022a}.  In particular, the functional connectome can shed new light onto neurological conditions such as schizophrenia and \gls{asd}, two conditions that alter brain function from healthy, neurotypical controls\cite{Woodward2015, Shi2017}.

A popular approach used to investigate brain function is \gls{fmri}, a non-invasive neuroimaging technique that measures blood flow through the brain over time\cite{Ogawa1990}.  An \gls{fmri} image is a complex spatio-temporal picture of the brain with voxels (volumetric pixels) describing the spatial location and a time series for each voxel describing the blood flow over time.  To reduce the spatial complexity, voxels can be collated into user-specified \glspl{roi}.  Functional connectomes can then be investigated through the Pearson correlation matrix between \glspl{roi}, known as the functional connectivity matrix.

One approach to investigating functional connectivity is using the \gls{pls} regression method.  Introduced by Wold (1975)\cite{Wold1975} for use in chemometrics, \gls{pls} is an extension of multivariate multiple regression to high-dimensional data that predicts the response data from a set of lower-dimensional latent variables constructed from the predictor data.  Popularised for \gls{fmri} by McIntosh \textit{et. al.} (1996)\cite{McIntosh1996}, \gls{pls} has been used to explore the relationships between \gls{fmri} data and either behavioural data, experimental designs, or seed region activation \citep{Krishnan2011}.  However, standard \gls{pls} ignores the structure of functional connectivity data -- functional connectivity matrices are correlation matrices and hence positive definite.

The space of $R\times R$ symmetric positive definite matrices -- which includes functional connectivity matrices -- forms a convex cone in $R(R+1)/2$-dimensional Euclidean space.  However, when considered with the affine invariant geometry\cite{Pennec2006}, the space of symmetric positive definite matrices becomes a complete Riemannian manifold with non-positive curvature.  By considering this non-linear geometry on symmetric positive definite matrices we can glean interesting new insights into functional connectivity (see Pennec \textit{et. al.} (2019)\cite{Pennec2019} and citations therein).

Here we propose an extension of the \gls{pls} model to allow Riemannian manifold response and predictor data, which we call \gls{rpls}. The \gls{rpls} model then allows us to predict from functional connectivity data while accounting for the intricate relationships enforced by the positive definite criteria.   To fit the \gls{rpls} model, we propose the \gls{tnipals} algorithm, which is related to  previously proposed applications of \gls{pls} for functional connectivity data in the literature\cite{Wong2018, Chu2020, Zhang2018a, Perez2013}.  We determine the optimal number of latent variables using cross validation.   To aid in interpretability of the high-dimensional functional connectivity data, we determine significant functional connections identified by \gls{rpls} using permutation tests on the \gls{vip} statistic\cite{Wold1993}, a popular measure of variable importance from standard \gls{pls}.

We apply \gls{rpls} to two datasets and two different \gls{roi} atlases to demonstrate its versatility.  First is the COBRE dataset\cite{Aine2017} which investigates differences in functional connectivity between health controls and patients with schizophrenia; we consider the \gls{fmri} in the COBRE dataset in the functional \gls{msdl} atlas\cite{Varoquaux2010}.  Second is the ABIDE dataset\cite{Craddock2013} which investigates differences in functional connectivity between typical health controls and subjects with \gls{asd}; we consider the ABIDE data in the \gls{aal} atlas \cite{Tzourio-Mazoyer2002}.

\section*{Results}


\subsubsection*{COBRE}

Ten-fold cross validation showed that $K = 2$ latent variables was the most parsimonious, within one standard error of the minimum \gls{rmse} ($K=3$).  When compared with Euclidean \gls{pls} using raw and Fisher-transformed correlations,  \gls{rpls} outperformed both methods across all metrics except for specificity in group prediction  (Table \ref{tab:data_results}) .  However, all three methods produced similar results for every metric.

\begin{figure}[tbph]
	\begin{subfigure}[b]{0.48\linewidth}
		\centering
		\includegraphics[width=\linewidth]{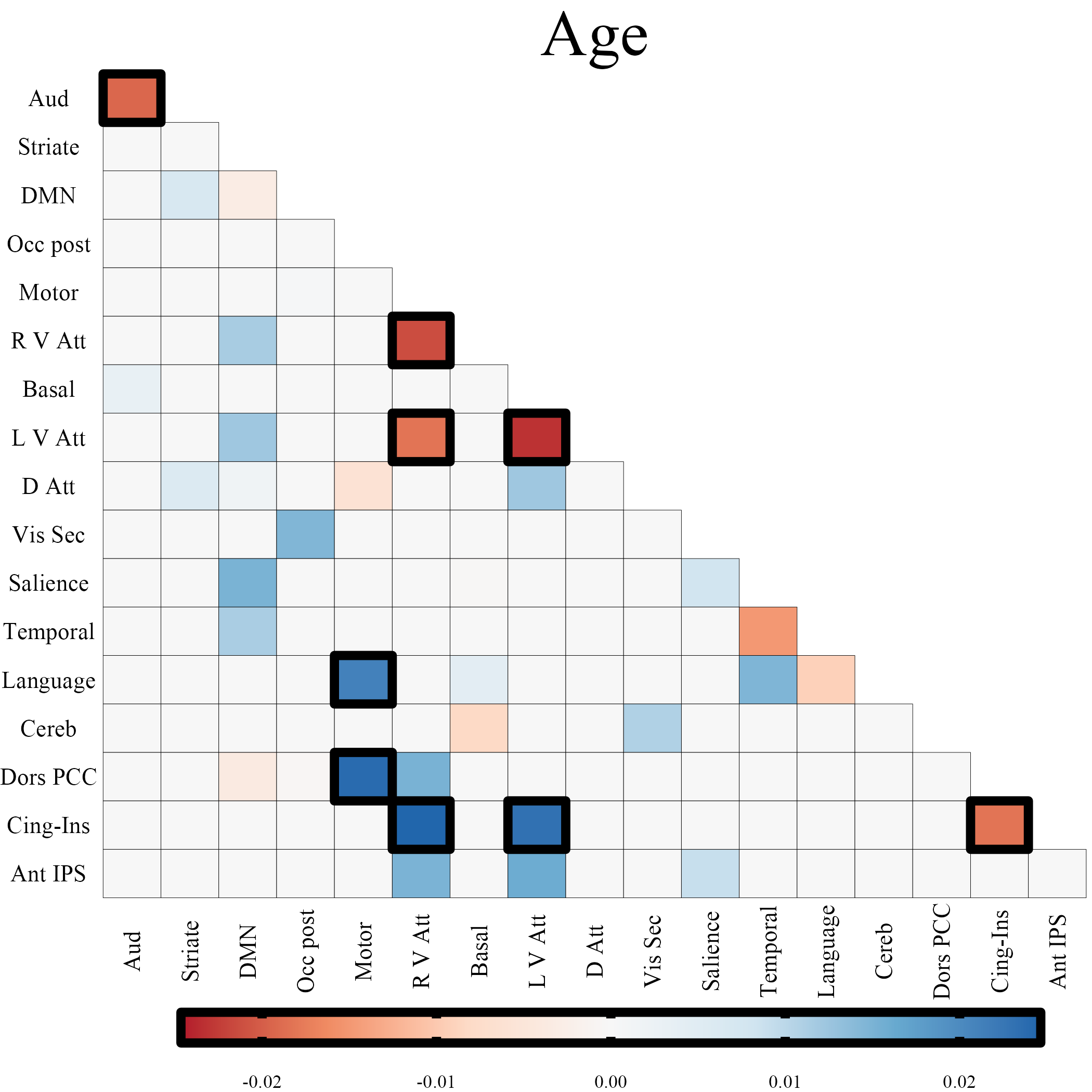}
	\end{subfigure}
	\begin{subfigure}[b]{0.48\linewidth}
		\centering
		\includegraphics[width=\linewidth]{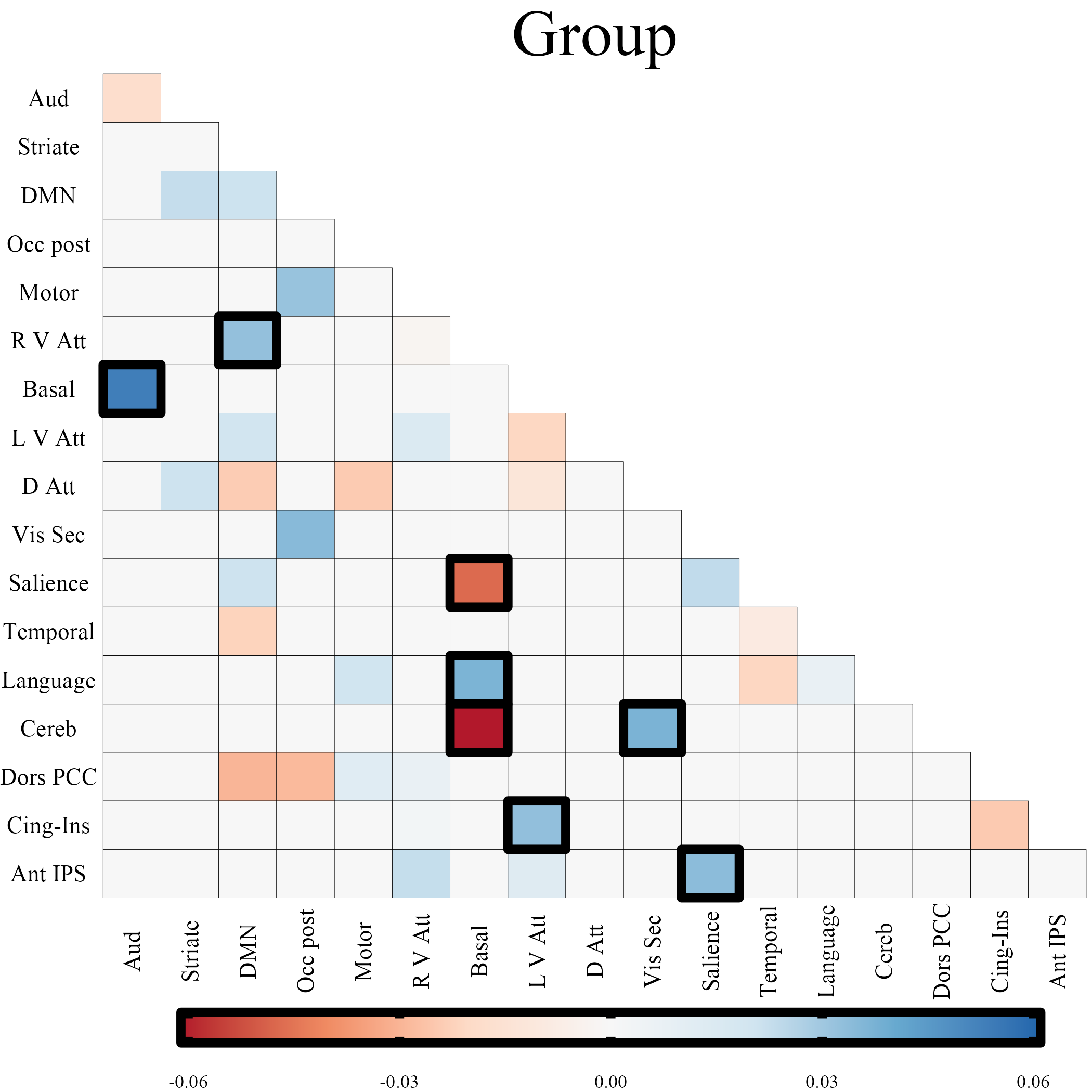}
	\end{subfigure}
	\centering
	\caption{}
	\label{fig:cobre_coefficients}
\end{figure}

A permutation test of the $\mathrm{VIP}$ statistic (Equation \ref{eqn:vip_original}) with $200$ permutations found  $45$ significant functional connections between \glspl{roi} as being predictive of age and subject group (Figure \ref{fig:cobre_coefficients}).  To aid interpretability, we have reduced the $39$ \glspl{roi} of the \gls{msdl} atlas into the $17$ resting state networks associated to the atlas \cite{Varoquaux2011} by taking the mean coefficient value within the \glspl{roi} of each network, as suggested by Wong \textit{et. al.} (2018) \cite{Wong2018}.

An increase in subject age tended towards a decrease of within-network connectivity (as measured by a mean decrease in functional connectivity within-networks) with particular emphasis on the auditory network, cingulate insula, and left and right ventral attention networks (Figure \ref{fig:cobre_coefficients} (left)).  Further, increased age was associated with an increase in between-network connectivity with focus on connectivity involving the cingulate insula and the motor network.

For subjects in the schizophrenic group, the basal ganglia exhibited both increased and decreased connectivity with other networks (Figure \ref{fig:cobre_coefficients} (right)).  In particular, there was a decrease in connectivity between the basal ganglia and the cerebellum and salience networks, whereas we observed an increase in connectivity between the basal ganglia and auditory and language networks for the schizophrenic group.  We also note that the default mode network was highly discriminatory for the schizophrenic group showing increased within-network connectivity and both increased and decreased between-network connectivity.

\subsubsection*{ABIDE}

Ten-fold cross validation found $K = 3$ latent variables was the most parsimonious, within one standard error of the minimum \gls{rmse} ($K=6$).  When compared with Euclidean \gls{pls} using the raw and Fisher-transformed correlations,  \gls{rpls} outperformed both methods across all metrics except for specificity in group classification (Table \ref{tab:data_results}).  In particular, the $R^2$ value for \gls{rpls} was substantially larger than the Euclidean methods.

\begin{figure}[tbph]
	\begin{subfigure}[b]{0.48\linewidth}
		\caption{}
		\centering
		\includegraphics[width=\linewidth]{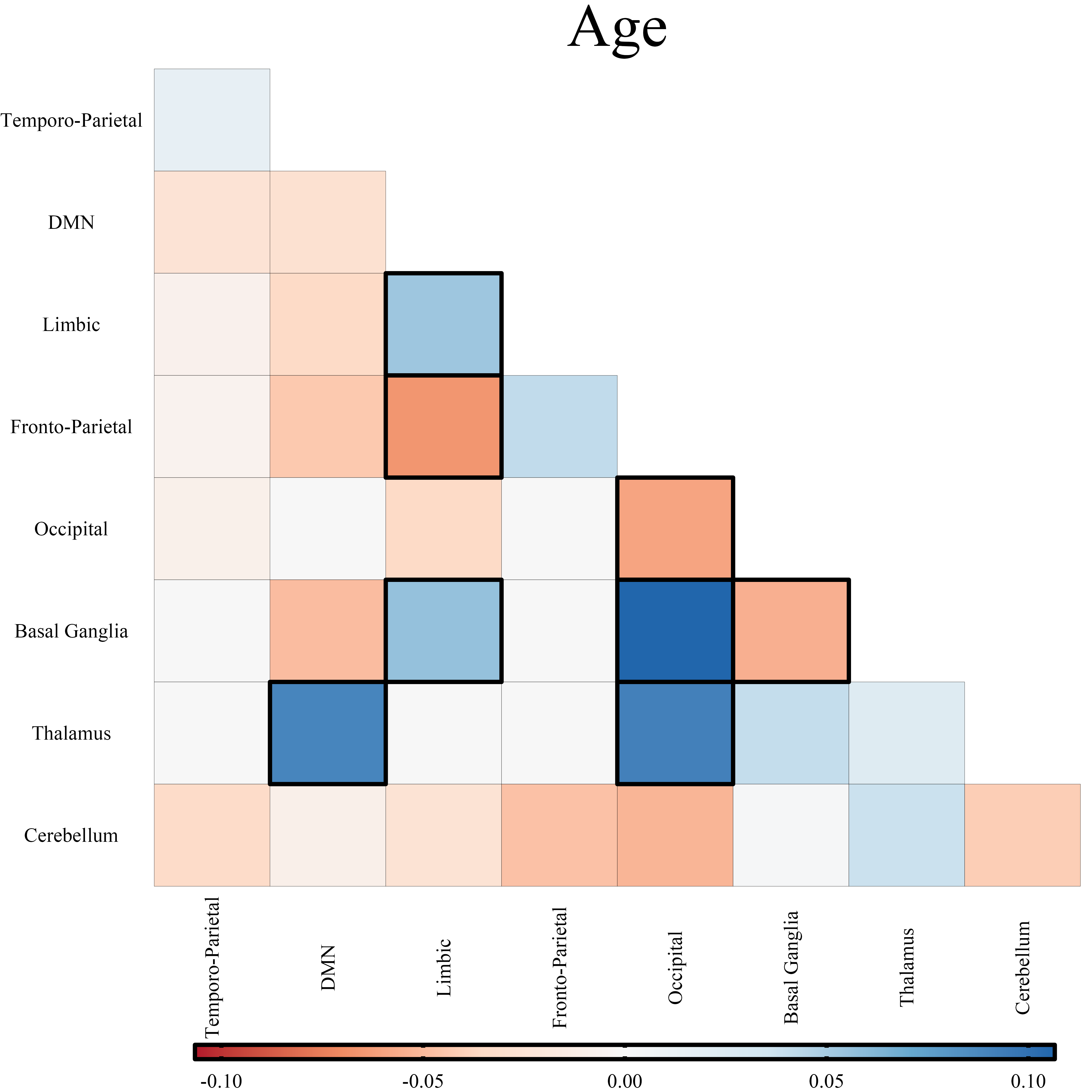}
	\end{subfigure}
	\begin{subfigure}[b]{0.48\linewidth}
				\caption{}
		\centering
		\includegraphics[width=\linewidth]{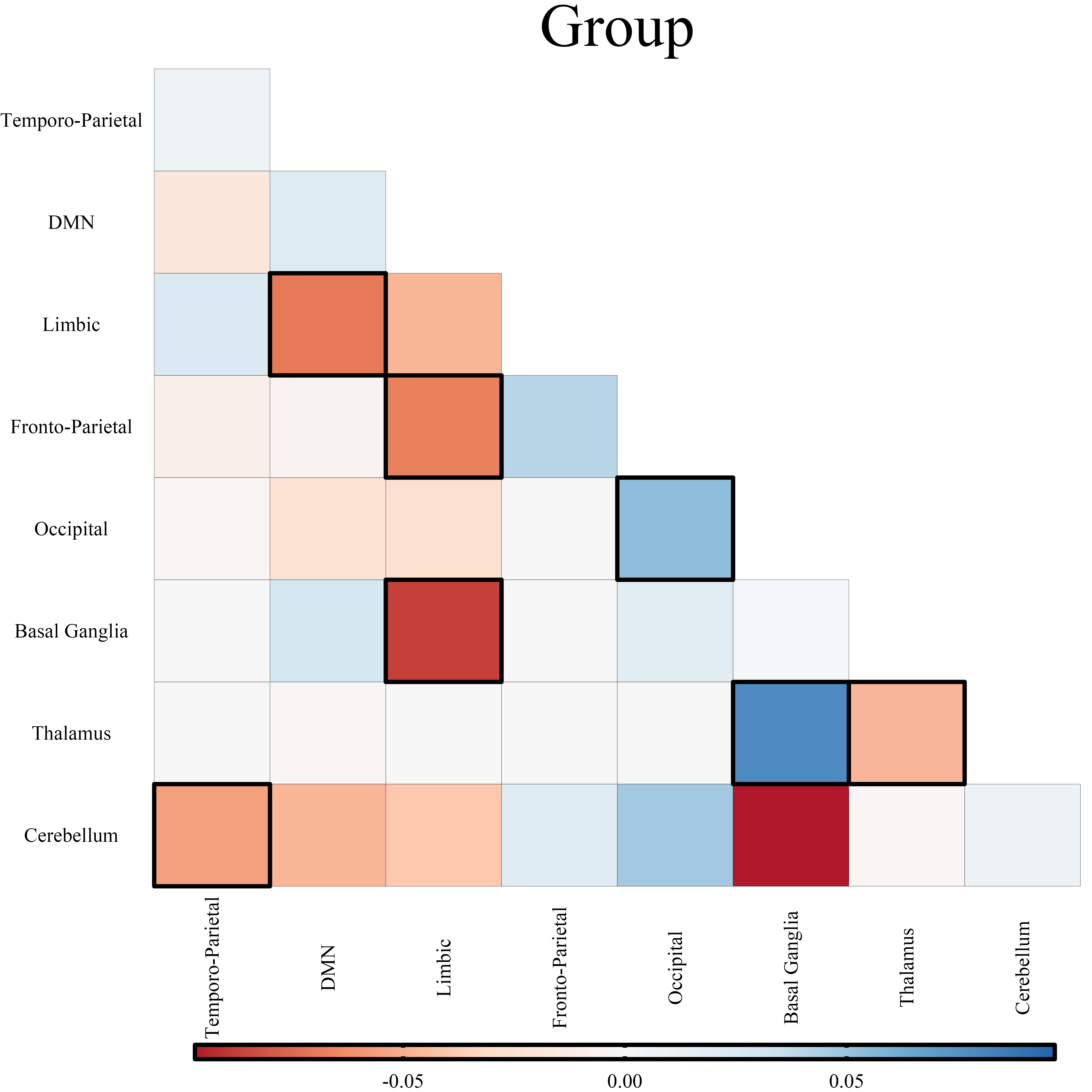}
	\end{subfigure}
	\begin{subfigure}[b]{0.48\linewidth}
		\caption{}
		\centering
		\includegraphics[width=\linewidth]{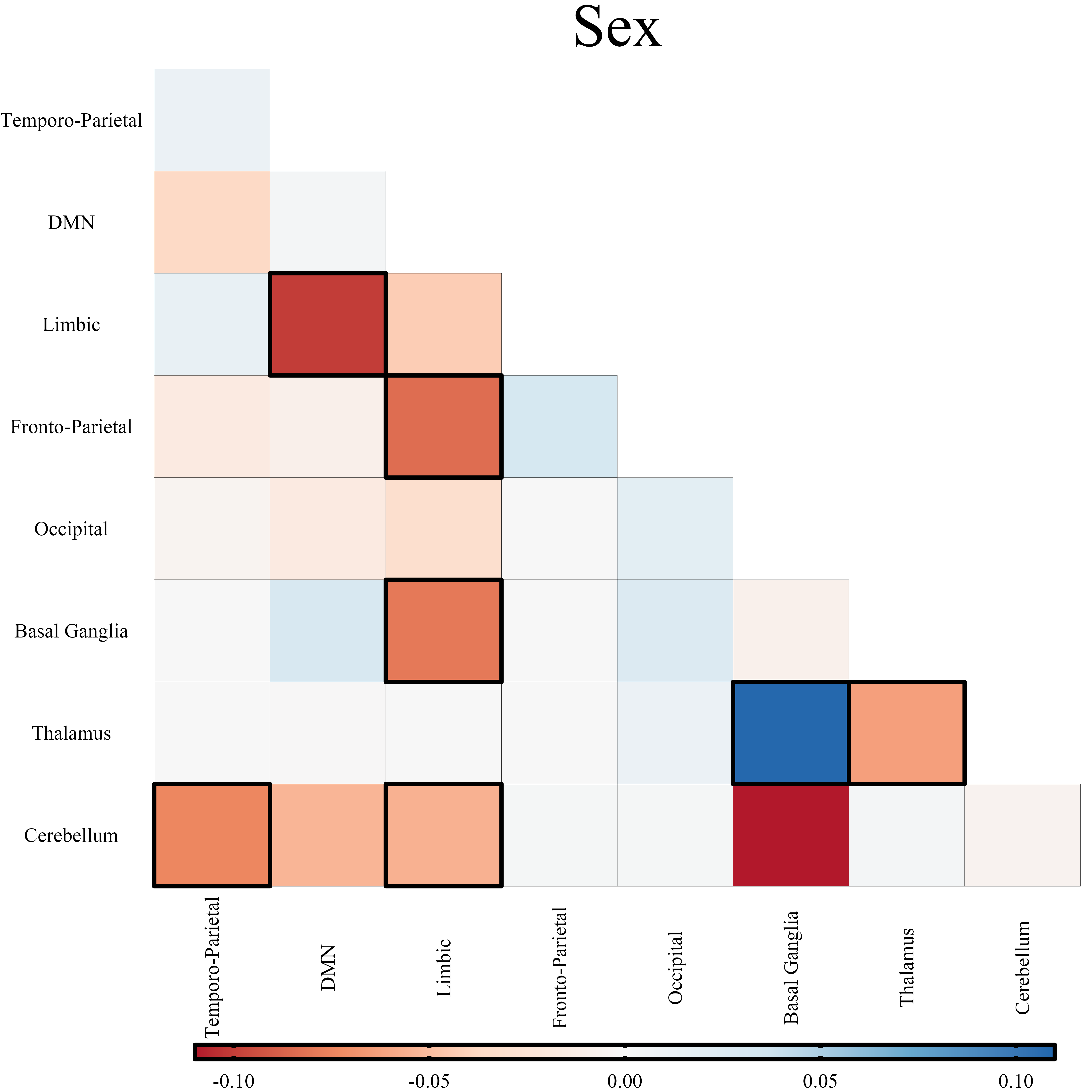}
	\end{subfigure}
	\begin{subfigure}[b]{0.48\linewidth}
		\caption{}
		\centering
		\includegraphics[width=\linewidth]{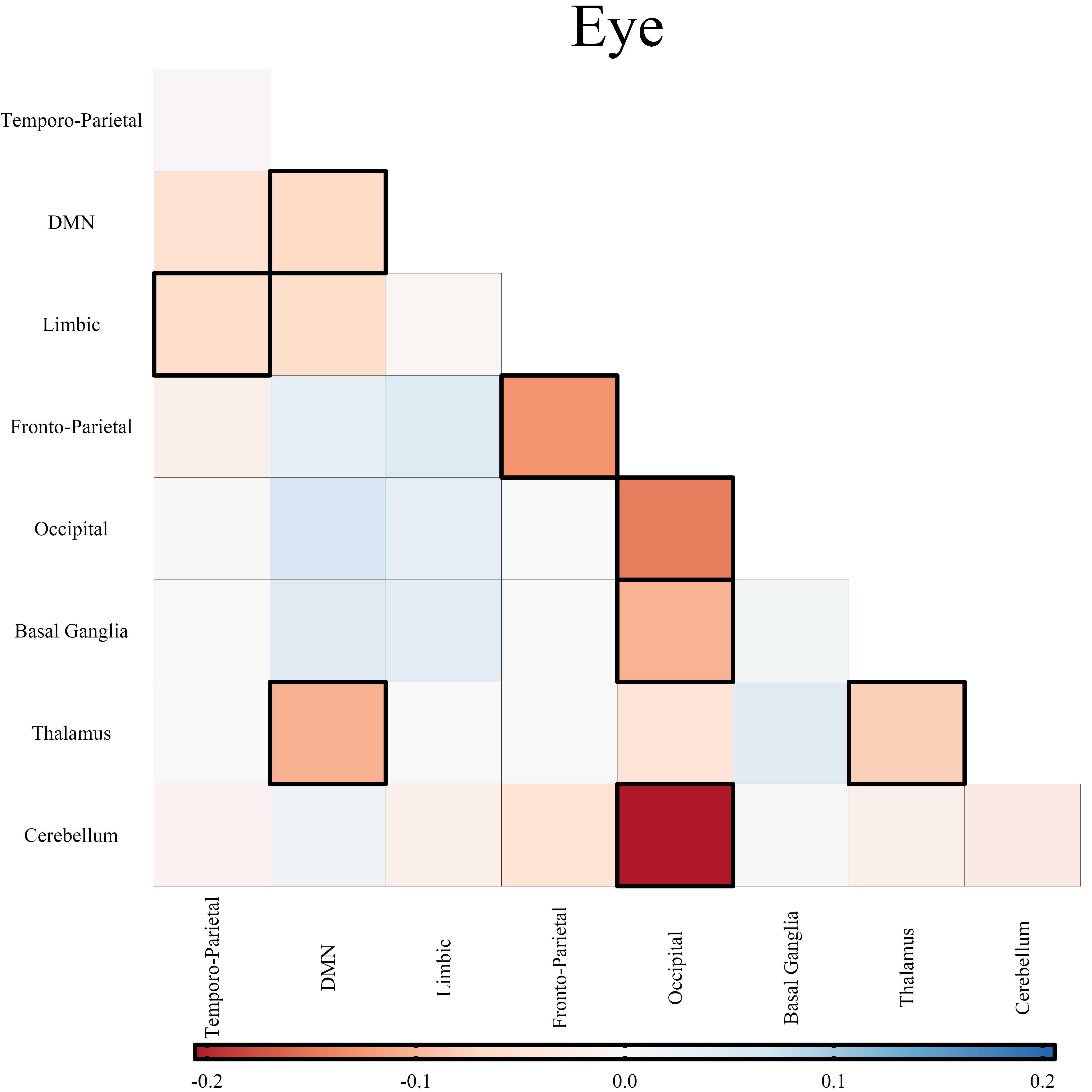}
	\end{subfigure}
	\centering
	\caption{}
	\label{fig:abide_coefficients}
\end{figure}

A permutation test of the  $\mathrm{VIP}$ statistic (Equation \ref{eqn:vip_original}) with $200$ permutations found $208$ significant functional connections between \glspl{roi} as being predictive of age, subject group, sex and eye status (Figure \ref{fig:abide_coefficients}).  We aid interpretability by associating the $116$ \glspl{roi} of the \gls{aal} atlas to the seven resting-state networks suggested by Parente and Colosimo (2020)\cite{Parente2020} and an eighth containing the cerebellum and vermis, which we call the cerebellum network.

In the ABIDE dataset, increased age was associated to both increased and decreased functional connectivity within resting-state networks (Figure \ref{fig:abide_coefficients} (a)).  Although we observed increased between-network connectivity for the thalamus and occipital networks, the cerebellum and default mode network exhibited decreased between-network connectivity with age.

For subjects with \gls{asd} we observed increased within-network connectivity with the exception of the limbic network and the thalamus  (Figure \ref{fig:abide_coefficients} (b)).  We also observed decreased between-network connectivity particularly for the cerebellum and the limbic networks.  We observed the same connectivity patterns for subject sex (Figure \ref{fig:abide_coefficients} (c))

For subjects with their eyes closed, our model suggests there was decreased within-network connectivity (Figure \ref{fig:abide_coefficients} (d)).  With the exception of the default mode network and the limbic network, we saw decreased between-network connectivity with particular emphasis on the occipital network.

\section*{Discussion}

The \gls{rpls} model has identified many functional connections associated to age, \gls{asd}, schizophrenia, sex, and eye status that are well represented in the literature.  In both the COBRE and ABIDE datasets, we identified the reduction of within-network connectivity with age that has been previously observed\citep{Varangis2019, Edde2021, Ferreira2016}, with exceptions in the temporo-parietal, fronto-parietal, limbic and thalamus networks in the ABIDE dataset and the salience network in the COBRE dataset, which all show an increase in connectivity with age. Further, both datasets exhibit the decreased connectivity with the default mode network, consistent with existing literature \citep{Tomasi2011, Pinero2014}.

For subjects with \gls{asd},  the decreased connectivity with the cerebellum\cite{Ramos2019} and the limbic\cite{Pascual2018} networks have been previously observed.  However, the decreased between-network connectivity suggested by \gls{rpls} is in contradiction with existing literature\citep{Assaf2010, Wong2018}; in particular, Wong \textit{et. al.} (2018)\cite{Wong2018} showed an increase in between-network connectivity associated to \gls{asd} on the full ABIDE dataset using logistic regression.  Also, observe that the connectivity for subject sex is highly correlated with the connectivity for the \gls{asd} group.  Although interactions between subject sex and \gls{asd} have been identified \citep{Smith2019}, we believe this highlights a possible limitation of \gls{rpls} and requires further investigation in future research.

The role of the basal ganglia in schizophrenic patients has been previously observed, particularly the decrease in connectivity between the salience network and the basal ganglia \citep{Zhang2019, Orliac2013} and the decreased connectivity between the cerebellum and basal ganglia \citep{Duan2015}.  Further, the connectivity patterns involving the default mode network have been previously reported in schizophrenic patients \citep{Karbasforoushan2012, Woodward2011, Dong2018, Yu2012, Wang2015}.

The results for eye status during scan are  also well represented in the literature.  The decreased within-network connectivity for the default mode network for patients with closed eyes has been previously reported by Yan \textit{et. al.} (2009)\cite{Yan2009}, and the increased between-network connectivity for the default mode network has recently been discussed by Han \textit{et. al.} (2023)\cite{Han2023}.  Further, the observed decrease in connectivity for the occipital network agrees with Agcaoglu \textit{et. al.} (2019)\cite{Agcaoglu2019}.

The use of the \gls{vip} statistic to  identify significant connections in functional connectivity has not been previously studied.  We have demonstrated that this statistic can identify many functional connections that have been addressed previously in the literature, but it is not without its limitations.  First, with our focus on generalising partial least squares to Riemannian manifolds, the \gls{vip} statistic does not take into account the Riemannian geometry we are considering.  This is mitigated by the tangent space approximation we are performing, which directly accounts for the geometry of the data, but further research could help better generalise the \gls{vip} statistic for \gls{rpls}.  Further, the \gls{vip} statistic associates the effects of a single predictor on the full multivariate response.  In situations like we consider here, this makes it difficult to determine which functional connections are associated to which outcome variable.  For example, the connectivity within the default mode network is deemed significant by the \gls{vip} statistic in the ABIDE dataset, but it is unclear whether this connectivity is significance for every outcome variable or a subset of them.  Work has been done to generalise the \gls{vip} statistic when the outcome variable is multivariate\cite{Ryan2023}, but further research is needed to investigate this generalisation.

These results suggest that \gls{rpls} can provide insight into the functional connectome and how it relates to subject phenotype data.   Further, due to the specification and generality of the \gls{rpls} model, this method is readily applicable to other imaging modalities, and in particular to multimodal imaging studies.  The application of \gls{rpls} to multimodal imaging studies is an area of future research that may help to us to understand the functional networks that make up the human connectome.

\section*{Methods}

\subsection*{Data}

The International Neuroimaging Data-Sharing Initiative (INDI) is an initiative set to encourage free open access to neuroimaging datasets from around the world.  We consider two datasets that are accessible as a part of the INDI.

\subsubsection*{COBRE \label{sec:cobre}}

The Center for Biomedical Research Excellence (COBRE)  have contributed structural and functional MRI images to the INDI that compare schizophrenic patients with healthy controls \cite{Aine2017}.  The data were collected with single-shot full \textit{k}-space echo-planar imaging with a TR of $2000$ milliseconds, matrix size of $64\times 64$ and $32$ slices (giving a voxel size of $3\times 3\times 4\, mm^3$).  These data were downloaded using the \textsc{Python} package \texttt{nilearn v 0.6.2}, and contains $146$ subjects (Control $= 74$), each with phenotype information on subject group and age; further information is available in Table S1 of the supplementary material.

The \gls{fmri} data were preprocessed using NIAK 0.17 under CentOS version 6.3 with Octave version 4.0.2 and the Minc toolkit version 0.3.18 \cite{Bellec2011}.  The data were subjected to nuisance regression where we removed six motion parameters, the frame-wise displacement, five slow-drift parameters, average parameters for white matter, lateral ventricles, and global signal, as well as 5 estimates for component based noise correction \cite{Behzadi2007}.  

For the COBRE dataset, we consider each \gls{fmri} in the \gls{msdl} atlas, a  functional \gls{roi} decomposition of $39$ nodes across $17$ resting state networks  \cite{Varoquaux2011}.  Time series were extracted for each \gls{roi} by taking the mean time series across the voxels in each region.

\subsubsection*{ABIDE \label{sec:abide}} 

The Autism Brain Imaging Data Exchange (ABIDE) is part of the Preprocessed Connectomes Project in INDI \cite{Craddock2013}.  The ABIDE data is a collection of preprocessed \gls{fmri} images from $16$ international imaging sites with 539 individuals diagnosed with \gls{asd} and 573 \gls{ntc}.  The ABIDE initiative provides data preprocessed under four separate standard pipelines, as well as options for band-pass filtering and global signal regression.  

Here we consider the $172$ subjects (\gls{ntc} = $98$) of the New York University imaging site.  We restrict to this site to reduce inter-site variation in imaging and because it is the largest individual imaging site.  The data were collected with a 3 Tesla Allegra MRI using echo-planar imaging with a TR of $2000$ milliseconds, matrix size of $64\times 64$ and $33$ slices (giving a voxel size of $3\times 3\times 4 mm^3$).  The \gls{fmri} data were downloaded using the \textsc{Python} package \texttt{nilearn v 0.6.2} preprocessed using the NIAK 0.7.1 pipeline \cite{Bellec2011}.  The data were subjected to: motion realignment; non-uniformity correction using the median volume; motion scrubbing; nuisance regression which removed the first principal component of 6 motion parameters, their squares, mean white matter and cerebrospinal fluid signals, and low frequency drifts measured by a discrete cosine basis with a 0.01 Hz high-pass cut-off; band-pass filtering and; global signal regression.  We consider the subjects preprocessed \gls{fmri} as well as subject group, age, sex, and eye status during scan (open or closed); further information is available in Table S2 of the supplementary material.

For the ABIDE dataset, we consider each \gls{fmri} in the \gls{aal} atlas\cite{Tzourio-Mazoyer2002}, an anatomical atlas of $116$ nodes across the brain.  Time series were extracted by taking the mean time series across the voxels in each \gls{roi}.

\subsection*{Partial least squares in Euclidean space}

\Gls{pls} is a predictive modelling technique that predicts a response matrix $\mat{Y}_{n \times q}$ from a set of predictors $\mat{X}_{n \times p}$. Originally introduced in the chemometrics literature by Wold (1975) \cite{Wold1975}, \gls{pls} has found application in bioinformatics \citep{Nguyen2002}, social sciences \citep{Hulland1999}, and neuroimaging \citep{Krishnan2011, McIntosh2004, Lin2003}; see Rosipal and Kr\"amer (2006)\cite{Rosipal2006} and citations therein for further examples.   As an extension of multivariate multiple regression, \gls{pls} has been shown to have better predictive accuracy than multivariate multiple regression when the standard regression assumptions are met \citep{Garthwaite1994}.   A further advantage of \gls{pls} is that it is effective when $q > n$ or $p > n$ since it performs prediction from lower dimensional latent variables, that is, \gls{pls} constructs a new set of predictor variables from $\mat{X}$ to predict $\mat{Y}$ \citep{Garthwaite1994}.  

Let $\mat{X}_{n \times p}$ and $\mat{Y}_{n \times q}$ be predictor and response matrices respectively.  Suppose $\mat{X}$ and $\mat{Y}$ are column centred, that is, suppose the means of each column of $\mat{X}$ and $\mat{Y}$ are $0$. \gls{pls} proposes the existence of $L \leq \min\{p, n\}$ latent variables such that $\mat{X}$ and $\mat{Y}$ decompose into a set of \textit{scores matrices} $\mat{T}_{n\times L}$ and $\mat{U}_{n \times L}$, and \textit{loadings matrices} $\mat{P}_{p\times L}$ and $\mat{Q}_{q\times L}$ with
\begin{align}
	\label{eqn:outer_relations_x}
	\mat{X} & = \mat{T} \mat{P}^T + \mat{E}\, ,\\
	\label{eqn:outer_relations_y}
	\mat{Y} & = \mat{U} \mat{Q}^T + \mat{F}\, , 
\end{align}
where $\mat{E}_{n\times p}$ and $\mat{F}_{n \times q}$ are error matrices, assumed to be a small as possible \citep{Geladi1986}, and the superscript $T$ denotes the matrix transpose.  Further, \gls{pls} assumes that there is a diagonal matrix $\mat{B}_{L \times L}$ with 
\begin{equation}
	\label{eqn:inner_relation}
	\mat{U} = \mat{T} \mat{B} + \mat{H}_{n\times L}\,, 
\end{equation}
where $\mat{H}$ is a matrix of residuals.  Equations \ref{eqn:outer_relations_x} and \ref{eqn:outer_relations_y} are called the \textit{outer relationships} while Equation \ref{eqn:inner_relation} defines the \textit{inner relationship} that connects $\mat{X}$ and $\mat{Y}$.  Combining the inner relationship and the outer relationship for $\mat{Y}$ gives 
\[
	\mat{Y}  =  \mat{T} \mat{B} \mat{Q}^T + (\mat{H}\mat{Q}^T  + \mat{F})\, , 
\]
which highlights that $\mat{Y}$ is a regression on the latent scores $\mat{T}$.  Further, notice that the error in $\mat{Y}$ is given by $\mat{H}\mat{Q}^T  + \mat{F}$, that is, error in $\mat{Y}$ is a combination of error inherent to the response data ($\mat{F}$) and error from the estimation of the inner relationship ($\mat{H}\mat{Q}^T$).  The inclusion of the residual matrix $\mat{H}$ can complicate discussion of the \gls{pls} method, so it is common to consider the estimated inner relationship
\(
	\mat{\hat{U}} \approx \mat{T} \mat{B}
\)
instead \citep{Hoskuldsson1988, Geladi1986}. 

Estimation of the \gls{pls} model (Equations \ref{eqn:outer_relations_x}--\ref{eqn:inner_relation}) is commonly done through the \gls{nipals} algorithm (Algorithm S1 in the supplementary material).  The inputs for the \gls{nipals} algorithm are the data matrices $\mat{X}$ and $\mat{Y}$ and the pre-specified number of latent variables $K$; noting that the true number of latent variables $L$ is unknown, the value $K$ can be chosen with methods such as cross validation. The \gls{nipals} algorithm outputs estimates  of the scores, loadings, and regression coefficients as well as matrices $\mat{W}_{p\times K}$ and $\mat{C}_{q\times K}$ known as the weights.  The weight matrices $\mat{W}$ and $\mat{C}$ are linear transformations of $\mat{P}$ and $\mat{Q}$ that more efficiently fit the \gls{pls} model and are defined within the \gls{nipals} algorithm; see the supplementary material for further information.  Using the results of the \gls{nipals} algorithm and Equations \ref{eqn:outer_relations_x}--\ref{eqn:inner_relation}, we can write
\[
    \mat{\hat{Y}} = \mat{X}\mat{\hat{\beta}}_{PLS}
\]
where
\begin{equation}
    \label{eqn:beta_pls}
    \mat{\hat{\beta}}_{PLS} = \mat{W}(\mat{P}^T\mat{W})^{-1}\mat{B}\mat{C}^T
\end{equation}
is the matrix of regression coefficients.  Using $\mat{\hat{\beta}}_{PLS}$ we see that \gls{pls} is a linear regression technique similar to ordinary least squares and ridge regression.

\subsubsection*{The VIP statistic}

To determine significant predictors of the response variables in the \gls{pls} model, we use the \gls{vip} statistic \citep{Wold1993}.  Suppose there are $p$ predictor variables, $q$ response variables, and $K$ latent variables extracted using \gls{nipals}.  Following Tennenhaus (1998)\cite{Tenenhaus1998}, the \gls{vip} statistic for the $j^{th}$ predictor variable is
\begin{equation}
	\label{eqn:vip_original}
	\mathrm{VIP}_j  = \sqrt{\frac{ p }{ \mathrm{Rd}(\mat{Y}, \mat{T}) }\sum\limits_{k = 1}^K \mathrm{Rd}(\mat{Y}, \vec{t_k}) \left( w_{jk} \right)^2}\, ,
\end{equation}
where $\vec{t}_k$ is the $k^{th}$ column of the score matrix $\mat{T}$, $w_{jk}$ is the $k^{th}$ weight for the $j^{th}$ predictor, 
$\mathrm{Rd}(\mat{Y}, \vec{t_k})  = \frac{1}{q} \sum\limits_{i = 1}^q \mathrm{cor}(\mat{Y}_i, \vec{t}_k)^2$, and $\mathrm{Rd}(\mat{Y}, \mat{T})	 =  \sum\limits_{k = 1}^K \mathrm{Rd}(\mat{Y}, \vec{t_k})$. The coefficient $\mathrm{cor}(\mat{Y}_i, \vec{t}_k)^2$ is the squared correlation between the $i^{th}$ response variable and the $k^{th}$ score.  The denominator $\mathrm{Rd}(\mat{Y}, \mat{T})$  in Equation \ref{eqn:vip_original} measures the proportion of variance in $\mat{Y}$ explained by $\mat{T}$, and the numerator $\mathrm{Rd}(\mat{Y}, \vec{t_k}) (w_{jk})^2$ measures the proportion of variance in $\mat{Y}$ described by the $k^{th}$ latent variable that is explained by the $j^{th}$ predictor \citep{Tran2014}.  Thus the \gls{vip} statistic measures the influence of each predictor on the explained variation in the model \citep{Galindo2014}.

Commonly, the \lq\lq greater than one\rq\rq \, rule is used to find predictors significantly associated with the response.  However, this rule is motivated by the mathematical properties of $\mathrm{VIP}_j$ rather than statistical properties \citep{Tran2014}.  Thus, we use a permutation test to determine significance of $\mathrm{VIP}_j$.  This is an alternative to Afanador \textit{et. al.} (2013)\cite{Afanador2013} who used $95\%$ jackknife confidence intervals to determine significance of \gls{vip}.

Specifically, for each predictor variable $j$ we permute the values $H$ times.  For each permutation $h =1, 2, \dots, H$ we refit the \gls{pls} model and calculate $\mathrm{VIP}_{j, h}$.  The \textit{P}-value for the $j^{th}$ $\mathrm{VIP}$ score is then
\begin{equation}
	\label{eqn:perm_p}
	\text{\textit{P}-value}_{j} = \frac{\#\, \left\{ \mathrm{VIP}_{j, h} > \mathrm{VIP}_{j} \right\}}{H}\, .
\end{equation}
For our data, the predictors are functional connectivity matrices. 
Thus, we know \textit{a priori} that the diagonal elements are uninformative since they are identically one.  Hence, if predictor $j$ describes a diagonal element we set $\text{\textit{P}-value}_{j} = 1$ for all $i$.  To account for the multiple comparisons problem, we adjust all \textit{P}-values using the false discovery rate\cite{Benjamini1995} and determine significance at a significance level of $\alpha = 0.05$.

\subsection*{Mathematical preliminaries}

\subsubsection*{Riemannian manifolds}

Intuitively speaking, a Riemannian manifold $M$ is a space where we can perform calculus, measure distances, and measure angles between tangent vectors.  More specifically, a smooth $d$-dimensional manifold $M$ is a connected, Hausdorff, second countable topological space that is covered by a set of coordinate charts $\{(U_i, \varphi_i:U_i\rightarrow \RR^d)\}_{i \in I}$, defined by some indexing set $I$, such that every point in $M$ belongs to a $U_i$ for some $i\in I$ and the intersection maps $\varphi_i \circ \varphi_j^{-1}$ are smooth as maps $\RR^d \rightarrow \RR^d$ for every $i, j \in I$.  These coordinate charts make the space $M$ \lq\lq locally Euclidean\rq\rq \,in the sense that every point has a neighbourhood that looks like Euclidean space.  Since concepts from differential calculus are local in nature, the construction of a smooth manifold allows us to perform calculus on these more general spaces.

An important concept in the study of manifolds is the tangent bundle $TM = \bigsqcup_{a \in M} T_{a}M$, where $T_{a}M$ is the tangent space at $a$.  The space $T_{a}M$ is defined as the set of equivalence classes of curves through $a$ such that $\gamma_1$ and $\gamma_2$ are equivalent if $\gamma_1'(0) = \gamma_2'(0)$, where the prime denotes the derivative.  Then $T_{a}M$ is a vector space that generalises the notion of vectors tangent to a surface to arbitrary smooth manifolds.

A \textit{Riemannian} manifold is a manifold $M$ together with a smooth map $g:M\times TM\times TM \rightarrow \RR$ such that  $g(a,\cdot, \cdot) = g_a:T_{a}M \times T_{a}M\rightarrow \RR$ is an inner product for every $a\in M$.  The Riemannian metric $g$ allows us to measure angles between tangent vectors and measure distances between points on the manifold $M$.  Further, $g$ is used to define geodesics (locally length minimising curves) $\gamma:[t_0, t_1]\rightarrow M$ between two points $a, b \in M$.  We only consider complete Riemannian manifolds here, which are spaces where every geodesic $\gamma$ has domain $\RR$.

Through geodesics we get the concepts of the Riemannian exponential and logarithm maps which allow us to smoothly move between the manifold and the tangent space.  The Riemannian exponential at a point $a\in M$ is a map $\Exp_{a}:T_{a}M\rightarrow M$ defined by $\Exp(a, \cdot)(\gamma) = \Exp_{a}(\gamma) = \gamma(1)$, where $\gamma$ is a geodesic such that $\gamma(0) = a$.  The Riemannian exponential is a smooth map that is locally diffeomorphic and hence has a local inverse denoted $\Log(a, \cdot) = \Log_{a} : M\rightarrow T_{a}M$ defined by $\Log_{a}(b) = \gamma'(0)$ where $\gamma(t)$ is a geodesic from $a$ to $b$.  For a point $b \in M$ close to $a$, we think of $\Log_{a}(b)$ as the shortest initial velocity vector based at $a$ pointing in the direction of $b$.  Further information on Riemannian manifolds can be found in the books by Lee (2011, 2012, 2018) \cite{Lee2011, Lee2012, Lee2018} or do Carmo (1992) \cite{Carmo1992}.  An accessible introduction for medical imaging can be found in the book edited by Pennec \textit{et. al. } (2019)\cite{Pennec2019}.

\subsubsection*{Fr\'echet mean}

To capture the centre of data on a manifold we consider the Fr\'{e}chet (or intrinsic) mean of data $X_1, X_2, \dots, X_n \in M$.  First, consider the Riemannian distance between two close points $X_1, X_2\in M$ defined by
\[
    d_g(X_1, X_2) = \left\|\Log_{X_1}(X_2)\right\|\,,
\]
where $\|\cdot\|$ is the norm in $T_{X_1}M$ induced by the Riemannian metric.  By generalising the sum of squared distances definition of the arithmetic mean, the Fr\'{e}chet mean\cite{Frechet1948} is given by
\[
    \mu_X = \argmin \sum_{i=1}^n d_g(X_i, \mu_X)^2\, .
\]
We solve for $\mu_X$ using gradient decent\cite{Pennec2019}; see Algorithm S2 in the supplementary material for further information.

\subsubsection*{The affine invariant geometry for symmetric positive definite matrices}

Let $GL_R\RR$ be the set of $R\times R$ real invertible matrices.  The set of symmetric positive definite matrices is defined by
\[
    S^+_R = \left\{ \mat{A}\in GL_R\RR : \mat{A}^T = \mat{A} \text{ and } \vec{v}^T \mat{A} \vec{v} > 0 \text{ for all } \vec{v} \in \RR^{R} \backslash \{\vec{0}\} \right\}\, ,
\]
where superscript $T$ denotes matrix transpose.  The set $S_R^+$ is a smooth manifold, which can be easily seen by embedding $S_R^+$ into $\RR^{R(R+1)/2}$ as a convex cone.  This construction shows that the tangent space at each $\mat{A} \in S_R^+$ is given by the set of symmetric $R\times R$ matrices.

However, $S_R^+$ has an interesting intrinsic geometry known as the affine-invariant geometry \cite{Pennec2006}.  Under the affine invariant geometry $S_R^+$ becomes a complete Hadamard manifold -- a Riemannian manifold of non-positive curvature where $\Exp_{\mat{A}}$ is a diffeomorphism for every $\mat{A} \in S_{R}^+$.  

The affine-invariant metric $g$ is defined by
\[
    g_{\mat{A}}(\mat{U}, \mat{V}) = \tr\left( \mat{U} \mat{A}^{-1} \mat{V} \mat{A}^{-1} \right)\, ,
\]
where $\mat{A} \in S_R^+$, $\mat{U}, \mat{V} \in T_{\mat{A}}S_R^+$, and $\tr$ denotes the trace operator.  Using $g$, we can calculate the Riemannian distance between $\mat{A}, \mat{B} \in S_R^+$ as
\[
    d_g(\mat{A}, \mat{B})^2 = \sum_{r = 1}^R \left(\log\left( \sigma_r\left(\mat{A}^{-1/2} \mat{B} \mat{A}^{-1/2} \right) \right)\right)^2\, ,
\]
where $\sigma_r\left(\mat{A}^{-1/2} \mat{B} \mat{A}^{-1/2} \right)$ are the eigenvalues of $\mat{A}^{-1/2} \mat{B} \mat{A}^{-1/2}$, $r= 1, 2, \dots, R$.  Further, letting $\mat{A}, \mat{B} \in S_R^+$ and $\mat{U} \in T_{\mat{A}}S_R^+$, we get
\[
    \Exp_{\mat{A}}(\mat{U}) = \mat{A}^{1/2} \Exp\left( \mat{A}^{-1/2}\mat{U}\mat{A}^{-1/2} \right) \mat{A}^{1/2}
\]
and 
\[
    \Log_{\mat{A}}(\mat{B}) = \mat{A}^{1/2} \Log\left( \mat{A}^{-1/2}\mat{B}\mat{A}^{-1/2} \right) \mat{A}^{1/2}\, ,
\]
where $\Exp$ and $\Log$ are the matrix exponential and logarithm respectively.  The Riemannian distance, exponential, and logarithm are essential in the definition and fitting of the \gls{rpls} model defined below.

\subsection*{Riemannian PLS}

Let $M$ and $N$ be complete Riemannian manifolds.  Let $X_1, X_2, \dots, X_n \in M$ and $Y_1, Y_2, \dots, Y_n \in N$, and let $\mu_X$ and $\mu_Y$ denote the respective Fr\'{e}chet means.  Let $L \leq \min\{ \dim(M), n \}$.     The \gls{rpls} model proposes the existence of loadings $\vec{p}_1, \vec{p}_2, \dots, \vec{p}_L \in T_{\mu_X}M$ and $\vec{q}_1, \vec{q}_2, \dots, \vec{q}_L \in T_{\mu_Y}N$ such that,  for each subject $i = 1, 2, \dots, n$, there are scores $t_{i1}, t_{i2}, \dots, t_{iL} \in \RR$ and $u_{i1}, u_{i2}, \dots, u_{iL} \in \RR$ with
\begin{align}
	\label{eqn:riemannian_outer_relations_x}
	X_i & =\Exp\left( \Exp_{\mu_X}\left( \sum_{l = 1}^L t_{il} \vec{p}_l\right), \vec{e}_i\right)\, , \\
	\label{eqn:riemannian_outer_relations_y}
	Y_i & = \Exp\left(\Exp_{\mu_Y}\left( \sum_{l = 1}^L u_{il} \vec{q}_l \right), \vec{f}_i\right)\, , \text{ and }\\
	\label{eqn:riemannian_inner_relations}
	\hat{u}_{il} &= \hat{\beta}_{0l}  + \hat{\beta}_{1l} t_{il} \text{ for all }l = 1, 2, \dots, L\text{ and }i = 1, 2, \dots, n\,,
\end{align}
where  $\vec{e}_i \in T_{\Exp_{\mu_X}\left( \sum_{l = 1}^L t_{il} \vec{p}_l\right)} M$ and $\vec{f}_i \in T_{\Exp_{\mu_Y}\left( \sum_{l = 1}^L u_{il} \vec{q}_l \right)}M$ are error vectors with $\|\vec{e}_i\|$, $\|\vec{f}_i\|$ small. Equations \ref{eqn:riemannian_outer_relations_x} and \ref{eqn:riemannian_outer_relations_y} are the \textit{outer relationships} for Riemannian data, and Equation \ref{eqn:riemannian_inner_relations} is the \textit{inner relationship} connecting our response and predictor.  Note that, since the Riemannian exponential map on Euclidean space is vector addition, if $M = \RR^p$ and $N = \RR^q$ the \gls{rpls} model (Equations \ref{eqn:riemannian_outer_relations_x}--\ref{eqn:riemannian_inner_relations}) reduce to the standard \gls{pls} model (Equations \ref{eqn:outer_relations_x}--\ref{eqn:inner_relation}).

One approach to fitting \gls{rpls} is by directly generalising \gls{nipals} (Algorithm S1) to Riemannian manifolds (see, for example, Ryan (2023)\cite{Ryan2023}), but this becomes computationally intensive and fails to converge for sample sizes above $20$.  Instead, we propose a tangent space approximation to fitting \gls{rpls} when our data is close to the Fr\'echet mean, similar to methods such as Riemannian canonical correlations analysis \cite{Kim2014} and principal geodesic analysis \cite{Fletcher2013}.

The \gls{tnipals} algorithm (Algorithm \ref{alg:tRNIPALS}) works by first linearising the manifold data in a neighbourhood of the Fr\'{e}chet mean using the Riemannian logarithm (see supplementary material for further information), and then applying the Euclidean \gls{nipals} algorithm to the linearised data which is now vector-valued.  Thus, \gls{tnipals} provides a combination of the simplicity and efficiency of Euclidean \gls{nipals} with the geometry of the Riemannian manifold.  

The \gls{tnipals} algorithm provides a more general approach to Wong \textit{et. al.}'s (2018) \citet{Wong2018} method for constructing predictors from functional connectivity matrices to predict \gls{asd} using \gls{pls} and logistic regression by considering a Euclidean response and symmetric positive definite predictor.   Similarly, the methods of Zhang and Liu (2018)\cite{Zhang2018a}  and Chu \textit{et. al.} (2020) \cite{Chu2020}  are also generalised by \gls{tnipals}.  The  \gls{tnipals} algorithm for \gls{rpls} is also closely related to the \gls{pls} method for symmetric positive definite  matrices offered by Perez and Gonzalez-Farias (2013) \cite{Perez2013}.  

\subsection*{Model fitting}

For each dataset we predict the phenotype information (age, group, sex, eye status) from the functional connectivity data using the \gls{rpls} model.  To deal with low-rank functional connectivity matrices in the ABIDE dataset (which are not positive definite), we consider regularised functional connectivity matrices $\mat{\tilde{F}} = \mat{F} + \mat{I}$ following Venkatesh \textit{et. al.} (2020) \cite{Venkatesh2020}, where $\mat{I}$ is the $116\times 116$ identity matrix.  For comparison, we also fit the standard \gls{pls} model using the upper triangle of the functional connectivity matrices as the predictors (raw correlations), as well as their Fisher transformed values (Fisher correlations).  

We determine the optimal number of latent variables through ten-fold cross validation using the \lq\lq within one standard error\rq\rq \, rule \cite{Hastie2009} when minimising the root mean square error.  Due to the interest in the COBRE and ABIDE datasets in investigating the differences between healthy controls and patients, we also present the group classification metrics of accuracy, sensitivity, and specificity.

To investigate the functional connectomes associated to each phenotype variable, we consider the regression coefficient matrix $\mat{\beta}_{PLS}$ (Equation \ref{eqn:beta_pls}) where the $i^{th}$ column represents the effect of the functional connectivity matrix on the $i^{th}$ response variable.  We visualise the columns of the matrix $\mat{\beta}_{PLS}$ as symmetric matrices in the tangent space of the Fr\'{e}chet mean for each dataset. All analysis was performed using \textsc{R} \cite{Rcore}.

\bibliography{main.bib}

\section*{Acknowledgements}


M.R. acknowledges the Australian Government Research Training Program Scholarship that funded this research.

\section*{Author contributions statement}

M.R. developed the methods and analysed the data with consultation from G.G., J.T. and M.H.  All authors reviewed the manuscript. 

\section*{Data availability statement}

The data and \textsc{R} package (\texttt{spdMatrices}) used to complete this work are available on GitHub (Matthew-Ryan1995/Riemannian-statistical-techniques-with-applications-in-fMRI).  The code to perform the analyses and generate the figures is also found on GitHub (Matthew-Ryan1995/R-PLS-for-functional-connectivity).

\section*{Additional information}


\subsection*{Competing interests}

The authors declare no competing interests.


\newpage
\subsection*{Figures}

\newpage
\subsection*{Figure captions}

\begin{enumerate}
    \item[\textbf{Figure 1: }] Significant regression coefficients as measured by $\mathrm{VIP}$ for the \gls{rpls} model on the COBRE dataset with $K = 2$ latent variables, visualised as symmetric matrices.  The regression coefficients have been averaged over the $17$ resting state networks of the \gls{msdl} atlas and show the coefficients for age (left) and subject group (right). The darker outlined boxes show the top $25\%$ influential regions  as measured by the absolute coefficient value within and between each network.
    \item[\textbf{Figure 2: }] Significant regression coefficients as measured by $\mathrm{VIP}$ for the \gls{rpls} model on the ABIDE dataset with $K = 3$ latent variables, visualised as symmetric matrices.  The regression coefficients have been averaged over the $7$ resting state networks identified by Parente and Colosimo (2020)\cite{Parente2020} as well as the cerebellum network comprising the cerebellum and vermis.  (a) shows the coefficients that predict age, (b) shows the coefficients that predict group, (c) shows the coefficients that predict sex, and (d) shows the coefficients that predict eye status.  The darker outlined boxes show the top $25\%$ influential regions  as measured by the absolute coefficient value within and between each network.
\end{enumerate}

\newpage
\subsection*{Tables}
\begin{table}[htbp]
	\caption{\label{tab:data_results}Mean (SE) 10-fold cross validation results for R-PLS on the COBRE and ABIDE datasets, and Euclidean PLS using the raw and Fisher transformed correlations.  The value $K$ represents the optimal number of latent variables for each model using the within one standard error rule.  The full model metrics are the multivariate $R^2$ and RMSE.  The group classification metrics look at the classification for subject group only.  R-PLS is the best model for both datasets over all model metrics, except for specificity (bold values).}
	\centering
	\begin{tabular}[t]{>{}l|>{}ccc}
		\toprule
		& Riemannian & Raw correlations  & Fisher correlations\\
		\midrule
            \addlinespace[0.3em]
        \multicolumn{4}{l}{\textbf{COBRE}}\\
		\em{\hspace{1.5em}\cellcolor{gray!6}K }&\cellcolor{gray!6} 2 & \cellcolor{gray!6}3 & \cellcolor{gray!6}3 \\
		\addlinespace[0.3em]
		\multicolumn{4}{l}{\hspace{.75em}\textbf{Full model metrics (SE)}}\\
		\em{\hspace{1.5em}\cellcolor{gray!6}{$R^2$}} & \textbf{\cellcolor{gray!6}{0.25 (0.035)}} & \cellcolor{gray!6}{0.23 (0.033)} & \cellcolor{gray!6}{0.23 (0.036)}\\
		\em{\hspace{1.5em}RMSE} & \textbf{1.20 (0.036)} & 1.21 (0.025) & 1.21 (0.026)\\
		\addlinespace[0.3em]
		\multicolumn{4}{l}{\textbf{\hspace{.75em}Group classification (SE)}}\\
		\em{\hspace{1.5em}\cellcolor{gray!6}{Accuracy}} & \textbf{\cellcolor{gray!6}{0.75 (0.045)}} & \cellcolor{gray!6}{0.73 (0.032)} & \cellcolor{gray!6}{0.74 (0.032)}\\
		\em{\hspace{1.5em}Sensitivity} & \textbf{0.81 (0.035)} & 0.70 (0.057) & 0.72 (0.055)\\
		\em{\hspace{1.5em}\cellcolor{gray!6}{Specificity}} & \cellcolor{gray!6}{0.69 (0.071)} & \textbf{\cellcolor{gray!6}{0.76 (0.048)}} & \textbf{\cellcolor{gray!6}{0.76 (0.048)}}\\
  \midrule
  		\addlinespace[0.3em]
		\multicolumn{4}{l}{\textbf{ABIDE}}\\
		\em{\hspace{1.5em}\cellcolor{gray!6}K }&\cellcolor{gray!6} 3 & \cellcolor{gray!6}3 & \cellcolor{gray!6}3 \\
		\addlinespace[0.3em]
		\multicolumn{4}{l}{\textbf{\hspace{.75em}Full model metrics (SE)}}\\
		\em{\hspace{1.5em}\cellcolor{gray!6}{$R^2$}} & \textbf{\cellcolor{gray!6}{0.15 (0.015)}} & \cellcolor{gray!6}{0.07 (0.016)} & \cellcolor{gray!6}{0.07 (0.016)}\\
		\em{\hspace{1.5em}RMSE} & \textbf{1.80 (0.051)} & 1.89 (0.059) & 1.89 (0.059)\\
		\addlinespace[0.3em]
		\multicolumn{4}{l}{\textbf{\hspace{.75em}Group classification (SE)}}\\
		\em{\hspace{1.5em}\cellcolor{gray!6}{Accuracy}} & \textbf{\cellcolor{gray!6}{0.58 (0.027)}} & \cellcolor{gray!6}{0.55 (0.032)} & \cellcolor{gray!6}{0.54 (0.032)}\\
		\em{\hspace{1.5em}Sensitivity} & \textbf{0.61 (0.058)} & 0.52 (0.064) & 0.51 (0.063)\\
		\em{\hspace{1.5em}\cellcolor{gray!6}{Specificity}} & {\cellcolor{gray!6}{0.53 (0.063)}} & \textbf{\cellcolor{gray!6}{0.58 (0.065)} }& \textbf{\cellcolor{gray!6}{0.58 (0.065)}}\\
		\bottomrule
	\end{tabular}
\end{table}

\newpage
\subsection*{Algorithms}

\begin{algorithm}
	\caption{Tangent non-linear iterative partial least squares.}\label{alg:tRNIPALS}
	\KwIn{Data $X_1, X_2, \dots, X_n$, $Y_1, Y_2, \dots, Y_n$, Desired number of components $K$.}
	\KwOut{PLS weights $\{\vec{w}_k\}_{k =  1}^K, \{\vec{c}_k\}_{k =  1}^K$, Scores $\{\vec{t}_k\}_{k =  1}^K, \{\vec{u}_k\}_{k =  1}^K$, Loadings $\{\vec{p}_k\}_{k =  1}^K$, and Regression coefficients $\{  \hat\beta_{1k} \}_{k =  1}^K$.}
	Calculate Fr\'{e}chet means $\mu_X, \mu_Y$ (Algorithm S2$^*$)\;
	Linearise the data by\;
	\qquad $\vec{x}_i \gets \Log_{\mu_X}{X_i}$\;
	\qquad $\vec{y}_i \gets \Log_{\mu_Y}{Y_i}$\;
	Map $\vec{x}_i$, $\vec{y}_i$ to Euclidean space via coordinates $\phi$ on $T_{\mu_X}M$ and $\psi$ on $T_{\mu_Y}M$\;
	Perform NIPALS (Algorithm S1$^*$) on $\{(\vec{x}_i, \vec{y}_i)\}$ to get weights $\{\vec{w}_k\}_{k =  1}^K, \{\vec{c}_k\}_{k =  1}^K$, scores $\{\vec{t}_k\}_{k =  1}^K, \{\vec{u}_k\}_{k =  1}^K$, loadings $\{\vec{p}_k\}_{k =  1}^K$, and regression coefficients $\{  \hat{\beta}_{1k} \}_{k =  1}^K$\;
	Map $\vec{w}_k, \vec{c}_k$ and $\vec{p}_k$ back to their appropriate tangent spaces using $\phi^{-1}$ and $\psi^{-1}$.
 \begin{flushright}
    \footnotesize{$^*$Found in the supplementary material.}
\end{flushright}
\end{algorithm}

\end{document}


\flushbottom
\maketitle

\thispagestyle{empty}

\section*{Summary statistics}

For the COBRE dataset, we have phenotype information on subject group, sex, handedness, and age (Table \ref{tab:cobre}).  In our analysis, we only consider subject group and age as these were previously found to be associated with the distribution of the functional connectivity matrices in this dataset\cite{Ryan2023}.

For the ABIDE dataset, we have phenotype information on subject group, sex, eye status during scan, age, and three measures of intellegence quotient (full scale, visual, and performance; Table \ref{tab:abide}).  In our analysis, we only consider subject group, sex, eye status and age as these were previously found to be associated with the distribution of the functional connectivity matrices in this dataset\cite{Ryan2023}.

\begin{table}[htbp]
	\caption{\label{tab:cobre}Summary statistics for the COBRE dataset.  Numeric values are presented as mean (sd), and count values are presented as $N$ ($\%$).}
	\centering
	\begin{tabular}[t]{l | rrr}
		\toprule
		& Patient & Control  & Total\\
		\midrule
            \addlinespace[0.3em]
            \cellcolor{gray!6}{Count} & \cellcolor{gray!6}{72 (49.3)} & \cellcolor{gray!6}{74 (50.7)} & \cellcolor{gray!6}{146 (100)} \\
		{Female} & {14 (19.4)} & {23 (31.1)} & {37 (25.3)} \\
  \cellcolor{gray!6}{Left-handed} & \cellcolor{gray!6}{10 (13.9)} & \cellcolor{gray!6}{1 (1.4)} & \cellcolor{gray!6}{11 (7.5)} \\
		{Age} & {38.2 (13.9)} & {35.8 (11.6)} & {37 (12.8)} \\
		\bottomrule
	\end{tabular}
 
\hspace{1.5em}\\
\hspace{1.5em}\\

 	\caption{\label{tab:abide}Summary statistics for the ABIDE dataset.  Numeric values are presented as mean (sd), and count values are presented as $N$ ($\%$).}
	\centering
	\begin{tabular}[t]{l | rrr}
		\toprule
		& Autism Spectrum Disorder & Neurotypical Controls  & Total\\
		\midrule
            \addlinespace[0.3em]
            \cellcolor{gray!6}{Count} & \cellcolor{gray!6}{75 (43)} & \cellcolor{gray!6}{98 (57)} & \cellcolor{gray!6}{172 (100)} \\
		{Female} & {10 (13.5)} & {26 (26.5)} & {36 (20.9)} \\
  \cellcolor{gray!6}{Closed eyes} & \cellcolor{gray!6}{10 (13.5)} & \cellcolor{gray!6}{15 (15.3)} & \cellcolor{gray!6}{25 (14.5)} \\
		{Age} & {14.8 (7.1)} & {15.8 (6.2)} & {15.3 (6.6)} \\
  \cellcolor{gray!6}{Full scale IQ} & \cellcolor{gray!6}{107.4 (16.4)} & \cellcolor{gray!6}{113.4 (13.1)} & \cellcolor{gray!6}{110.8 (14.9)} \\
  {Visual IQ} & {105.1 (15.9)} & {113.2 (12.6)} & {109.7 (14.6)} \\
  \cellcolor{gray!6}{Performance IQ} & \cellcolor{gray!6}{108.6 (17.3)} & \cellcolor{gray!6}{110.4 (13.6)} & \cellcolor{gray!6}{109.6 (15.2)} \\
		\bottomrule
	\end{tabular}
\end{table}

\newpage

\section*{NIPALS}

Let $\mat{X}_{n \times p}$ and $\mat{Y}_{n \times q}$ be column centred predictor and response matrices respectively.  Consider the \gls{pls} model
\begin{align*}
	\mat{X} & = \mat{T} \mat{P}^T + \mat{E}\, ,\\
	\mat{Y} & = \mat{U} \mat{Q}^T + \mat{F}\, \\
        \mat{U} &= \mat{T} \mat{B} + \mat{H}_{n\times L}\,.
\end{align*}
The scores $\mat{T}, \mat{U}$, loadings $\mat{P}, \mat{Q}$, and regression matrix $\mat{B}$ can be iteratively calculated using the \gls{nipals} algorithm (Algorithm \ref{alg:NIPALS}).  Further, Algorithm \ref{alg:NIPALS} returns the weights matrices $\mat{W}$ and $\mat{C}$ which allow \gls{pls} to more efficiently predict on new data.

\begin{algorithm*}
	\caption{Non-linear iterative partial least squares (NIPALS)} \label{alg:NIPALS}
	\KwIn{Predictor matrix $\mat{X}$, Response matrix $\mat{Y}$, number of components $K$}
	\KwOut{Weights $\mat{W}, \mat{C}$, Scores $\mat{T}, \mat{U}$, Loadings $\mat{P}, \mat{Q}$, Regression matrix $\mat{B}$}
	$\mat{X}^1\gets \mat{X}$\;
	$\mat{Y}^1\gets \mat{Y}$\;
	\For{k = 1, 2, \dots, K}{
		\textbf{Calculate the scores and weights}\;
		$\vec{u} \gets \mat{Y}^k[, 1]$\;
		\Repeat{convergence}{
			$\vec{w} \gets( \mat{X}^k)^T \vec{u}/\vec{u}^T\vec{u}$\;
			$\vec{w} \gets \vec{w}/\|\vec{w}\|$\;
			$\vec{t} \gets \mat{X}^k\vec{w}$\;
			$\vec{c} \gets (\mat{Y}^k)^T \vec{t}/\vec{t}^T\vec{t}$\;
			$\vec{c} \gets \vec{c}/\|\vec{c}\|$\;
			$\vec{u} \gets \mat{Y}^k\vec{c}$\;
		}	
		\textbf{Calculate the loadings}\;
		\quad $\mat{X}^k$-loadings: $\vec{p} \gets (\mat{X}^k)^T \vec{t} / \vec{t}^T \vec{t}$\;
		\quad $\mat{Y}^k$-loadings: $\vec{q} \gets (\mat{Y}^k)^T \vec{u} / \vec{u}^T \vec{u}$\;
		\textbf{The regression step}\;
		\quad $b_k \gets \vec{u}^T \vec{t} / \vec{t}^T \vec{t}$\;
		\textbf{The deflation step}\;
		\quad $\mat{X}^{k+1} \gets \mat{X}^k - \vec{t} \vec{p}^T$\;
		\quad $\mat{Y}^{k+1} \gets \mat{Y}^k - b_k \vec{t} \vec{c}^T$\;
		Save\;
		\quad $\mat{W}[, k] = \vec{w}$\;
		\quad $\mat{C}[, k] = \vec{c}$\;
		\quad $\mat{T}[, k] = \vec{t}$\;
		\quad $\mat{U}[, k] = \vec{u}$\;
		\quad $\mat{P}[, k] = \vec{p}$\;
		\quad $\mat{Q}[, k] = \vec{q}$\;
		\quad $\mat{B}[k, k] = b_k$\;
	}
\end{algorithm*}

\newpage

\section*{The Fr\'echet mean}

Let $M$ be a Riemannian manifold and $X_1, X_2, \dots, X_n \in M$ be data.  Denote the Riemannian distance function by $d_g$ and recall the Fr\'echet mean is given by
\[
    \mu_X = \argmin \sum_{i=1}^n d_g(X_i, \mu_X)^2\, .
\]
Following do Carmo (1992) \cite{Carmo1992} the gradient of $d_g$ is given by
\[
    \mathrm{grad}_x d_g(y, x) = -2\mathrm{Log}(x, y)\, .
\]
Gradient descent can then be applied to calculate the Fr\'echet mean using Algorithm \ref{alg:frechet_mean}.
\begin{algorithm}
	\caption{\label{alg:frechet_mean}Gradient descent to calculate the Fr\'echet mean on the Riemannian manifold $M$.  This is Algorithm 2.1 from Pennec \textit{et. al.} (2019)\cite{Pennec2019} with an adapted step size $\tau$.}
	\KwIn{Data $\vec{y}_1, \vec{y}_2, \dots, \vec{y}_n \in M$, tolerance $\varepsilon$, step size $\tau$}
	\KwOut{$\mu_Y \in M$, the Fr\'{e}chet mean}
	Set $\mu_Y^{(0)} = \vec{y}_1$\;
	\While{$\varepsilon^{(k)} > \varepsilon$}{
		$\vec{v} = \frac{\tau}{n} \sum\limits_{i  =1}^n \Log_{\mu_Y^{(k)}}(\vec{y}_i)$\;
		\uIf{$\left\| \vec{v} \right\|_{\mu_Y^{(k)}} > \varepsilon^{(k)}$}{
		$\tau = \tau/2$\;
		$\mu_Y^{(k + 1)} = \mu_Y^{(k)}$
	}
	\Else{
		$\mu_Y^{(k + 1)} = \Exp_{\mu_Y^{(k)}}(\vec{v})$\;
		$\varepsilon^{(k + 1)} = \left\| \vec{v} \right\|_{\mu_Y^{(k)}}$
			}
	}
\end{algorithm}

\newpage

\section*{Linearising functional connectivity matrices in the affine invariant geometry}

To fit the \gls{rpls} model using \gls{tnipals}, we first need to linearise the manifold data at the Fr\'echet mean to get a vector representation, on which we can perform Euclidean \gls{nipals} (Algorithm \ref{alg:NIPALS}).  We discuss this further for the symmetric positive definite matrices $S_R^+$ equipped with the affine invariant metric following the exposition in Pennec \textit{et. al.} (2019)\cite{Pennec2019}.

Let $X_1, X_2, \dots, X_n \in S_R^+$ and let $\mu_X$ denote their Fr\'echet mean.  We first define the function $\mathrm{Vec}:T_{\mat{I}}S_R^+ \rightarrow \RR^{R(R+1)/2}$. 
 Recall that an element $\mat{U} \in T_{\mat{I}}S_R^+$ is a symmetrix matrix, which we write
\[
    \mat{U} = \begin{bmatrix}
        u_{11} & u_{12} & \dots & u_{1R}\\
        u_{12} & u_{22} & \dots & u_{2R}\\
        \vdots & \vdots & \ddots & \vdots \\
        u_{1R} & u_{2R} & \dots & u_{RR}\\
    \end{bmatrix}\, .
\]
Define
\[
    \mathrm{Vec}(\mat{U}) = \left( u_{11}, u_{22}, \dots, u_{RR}, \sqrt{2}u_{12}, \sqrt{2} u_{13}, \dots, \sqrt{2}u_{RR} \right)^T\, .
\]
Then $\mathrm{Vec}$ is a smooth isometry between $T_{\mat{I}}S_R^+$ and $\RR^{R(R+1)/2}$.  Using the affine invariant geometry, $\mathrm{Vec}$ can be extended to $\mu_X$ by
\[
    \mathrm{Vec}_{\mu_X}(\mat{U}) = \mathrm{Vec}\left( \mu_X^{-1/2} U \mu_X^{-1/2} \right)\, ,
\]
where $\mat{U} \in T_{\mu_X}S_R^+$ and $\mu_X^{-1/2}$ is the inverse of the symmetric square root of $\mu_X$.  Recalling that
\[
    \Log_{\mat{\mu}_X}(\mat{X}_i) = \mat{\mu}_X^{1/2} \Log\left( \mat{\mu}_X^{-1/2}\mat{X}_i\mat{\mu}_X^{-1/2} \right) \mat{\mu}_X^{1/2}\, ,
\]
takes $\mat{X}_i$ to $T_{\mu_X} S_R^+$, we can linearise our data by mapping
\[
    \mat{X}_i \mapsto \mathrm{Vec}_{\mu_X}\left(\Log_{\mat{\mu}_X}(\mat{X}_i)\right)\, .
\]
This linearisation can be visualised with the following commutative diagram.

\[
\begin{tikzcd}
    \RR^{R(R+1)/2} & & \\
    & T_{\mat{I}}S_R^+ \arrow{lu}{\mathrm{Vec}} & T_{\mu_X} S^+_R \arrow[bend right = 45, swap, dotted]{llu}{\mathrm{Vec}_{\mu_X}} \arrow[swap]{l}\\
    &  & S_R^+ \arrow[swap]{u}{\Log_{\mu_X}} \arrow[bend left = 45, dotted]{lluu}{\mathrm{Vec}_{\mu_X}\circ \Log_{\mu_X}}
\end{tikzcd}\, .
\]

\bibliography{main.bib}